**Incoherent dark excitons extend Floquet-like physics across time and momentum**

Xing Zhu[1], Jacques Hawecker[1], Enrico Perfetto[2,3], Vivek Pareek[1], David R. Bacon[1], Harley Suchiang[1], Nanami Tomoda[1], Amrita Mitra[1], Takumi Fukuda[1], Sebastien Roux[1], Ouri Karni[4], Jack Featherstone[1], Kenji Watanabe[5], Takashi Taniguchi[6], Michael K. L. Man[1], Julien Madéo[1], Felipe H. da Jornada[7,8], Gianluca Stefanucci[2,3], Keshav M. Dani[1*]

*Corresponding author: kmdani@oist.jp

[1]Femtosecond Spectroscopy Unit, Okinawa Institute of Science and Technology Graduate University, Onna, Okinawa, Japan 904-0495
[2]Dipartimento di Fisica, Universit`a di Roma Tor Vergata, Via della Ricerca Scientifica 1, 00133 Rome, Italy
[3]INFN, Sezione di Roma Tor Vergata, Via della Ricerca Scientifica 1, 00133 Rome, Italy
[4]Physics & Informatics Laboratories, NTT Research Inc., 940 Stewart Dr, Sunnyvale, CA 94085, USA.
[5]Research Center for Electronic and Optical Materials, National Institute for Materials Science, 1-1 Namiki, Tsukuba 305-0044, Japan
[6]Research Center for Materials Nanoarchitectonics, National Institute for Materials Science, 1-1 Namiki, Tsukuba 305-0044, Japan
[7]Department of Materials Science and Engineering, Stanford University, Stanford, CA, USA.
[8]Stanford PULSE Institute, SLAC National Accelerator Laboratory, Menlo Park, CA, USA

**Floquet phenomena — where a coherent, time-periodic drive hybridizes the conduction band with energy-replicated copies of the valence band — have been demonstrated with intense optical fields and, more recently, other bosonic excitations. However, within mean-field approximations and the current understanding of Floquet phenomena, these effects survive only for the coherence time of the drive, limiting practical use and current demonstrations to the sub-picosecond regime. Here, consistent with recent theoretical predictions beyond the mean-field picture, using time- and angle-resolved photoemission spectroscopy in $WSe_2/MoS_2$ heterostructures, we show that incoherent, dense populations of dark interlayer excitons hybridize the conduction band with energy-replicated copies of the valence band, mimicking Floquet-like phenomena — now persisting for nearly a nanosecond. With increasing interlayer exciton density, the valence band and its replica evolve from an inverted-parabolic to a Mexican-hat dispersion, a signature of the predicted non-zero static coupling that hybridizes the bands and scales linearly with the exciton density. By tuning twist angle, we further show that dark excitons with net momentum create replicas that hybridize with bands offset not just in energy but also in momentum, providing another powerful degree of freedom beyond conventional Floquet engineering. These results establish a robust, long-lived route to engineering electronic states, beyond a coherent, zero-momentum drive.**

## Introduction

Strong optical excitation can qualitatively reshape the quasiparticle spectrum of a semiconductor[1–7]. Resonant pump pulses tuned to the lowest bright exciton generate a self-sustained time-dependent macroscopic polarization, whose lifetime defines the excitonic coherence time. The polarization couples valence and conduction bands (VBs and CBs) and the resulting hybridization drives a pronounced reconstruction of the quasiparticle bands, evolving from a nearly parabolic valence-band dispersion at low excitation density to the characteristic Mexican-hat-like profile at high density[8,9]. This evolution can also be viewed as the nonequilibrium analogue of the Bose–Einstein-condensate (BEC) to Bardeen–Cooper–Schrieffer (BCS) crossover of an excitonic insulator, with the macroscopic polarization playing the role of the excitonic order parameter.

In a screened mean-field description, the polarization does not decay after the optical drive is removed. Because it oscillates at a well-defined frequency set by the exciton energy, the same many-body physics admits an equivalent Bloch–Floquet description[10,11]. In this picture, replicas of the valence and conduction bands appear, shifted in energy by the exciton energy. Very recently, time- and angle-resolved photoemission spectroscopy (TR-ARPES) experiments revealed a replica of the VB at the exciton energy, and a hybridization between VB and CB consistent with theoretical predictions[12]. These measurements were performed at early pump–probe delays, below approximately 200 fs, within the expected excitonic coherence time.

In a real system, excitons are expected to undergo collisions and eventually lose coherence. Within the mean-field description, this loss of coherence leads to a decay in the amplitude of the VB-CB coupling and, with it, the associated band reconstruction disappears. In contrast, recent theoretical work[13] went beyond the mean-field approximation to reveal a previously unexpected static non-zero coupling between VB and CB even after the coherence of the excitons goes to zero. In their experiments, Pareek, et al[12] observed a rapid decay of the valence band reconstruction, consistent with the coherence time of excitons and with existing mean-field theories. Nonetheless, the measured population decay of the carriers was also rapid, making it challenging to unambiguously distinguish between the coherent and incoherent regimes. The fate of the spectral reconstruction upon entering the incoherent regime thus remains an open question, motivating further experimental and theoretical work. The presence of such a long-lasting, incoherent coupling would open up the tantalizing possibility of establishing a persistent Floquet-like coupling between VB and CB long after the excitonic coherence/polarization has decayed.

Here using time- and angle-resolved photoemission spectroscopy (TR-ARPES) on twisted bilayer $MoS_2/WSe_2$ heterostructures, we clearly separate the coherent and incoherent regime of excitons and observe that incoherent and dense populations of dark interlayer excitons (ILX) mimic Floquet-like effects. We show that a dense population of ILX produces band hybridization transforming the parabolic dispersion of the top valence band into a Mexican-hat-like dispersion in energy-momentum space, as seen previously in the coherent limit of exciton-driven Floquet effects[12]. The Floquet-like features induced by incoherent dark excitons survive for nearly a nanosecond, orders of magnitude beyond the lifetime of coherently dressed Floquet states. By increasing exciton density, we find that the hybridization energy is linearly proportional to the population of the incoherent ensemble of the excitons, consistent with the theoretical framework of Ref. [13]. Furthermore, we also demonstrate controllable hybridization of bands located at different momenta of the band structure, enabled by adjusting the heterostructure twist angle and thus the ILX finite momentum driving the effect. Overall, our results show long-lived Floquet-like states with momentum as a new degree of freedom beyond conventional Floquet engineering (Fig. 1a) to explore novel phases and topologies.

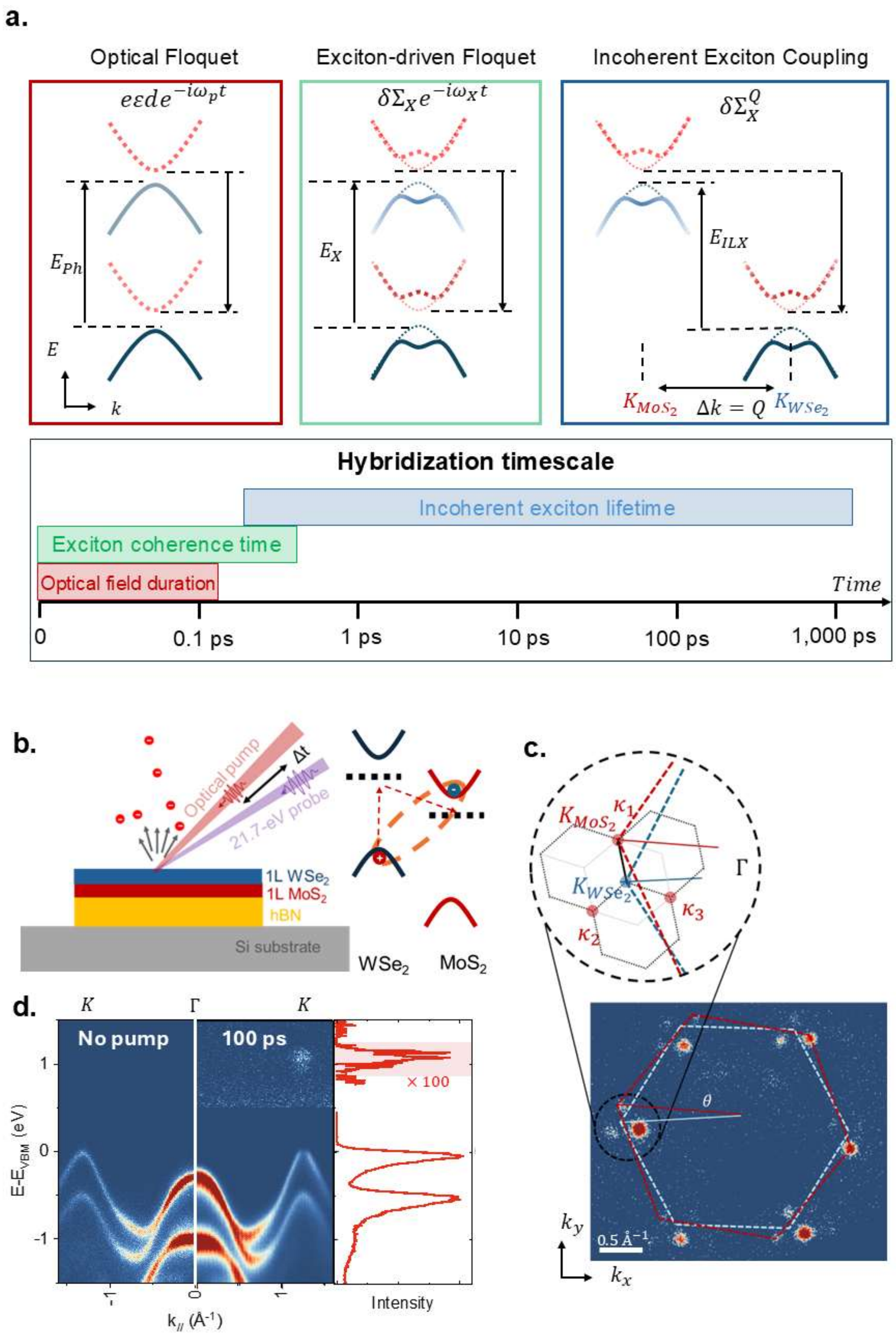

**Figure 1. Experimental study of electronic structure with high-density interlayer excitons in heterostructure $WSe_2/MoS_2$. a**, Comparison of optical Floquet, exciton-driven Floquet and incoherent static coupling driven by the dark excitons. For interlayer excitons, the incoherent static coupling could last for much longer time scale. **b**, Schematics of the TR-ARPES measurements on twisted bilayer $WSe_2/MoS_2$ heterostructure. The optical excitation at 1.72eV creates intralayer excitons in $WSe_2$ and then in quasi-equilibrium, an ensemble of incoherent dark ILXs forms with the electrons confined in $MoS_2$ layer and the holes in $WSe_2$ layer. **c**, The momentum distribution of the long-lived excited states around 1.1 eV. The experimental distribution of the electron appears at the three corners of miniBZ with momentum offset to the K/K' valleys in $WSe_2$. Schematics of the hexagonal Brillouin Zone of the $MoS_2$ (red) and $WSe_2$ (blue) show the relative twist angle $\theta$ between two layers. **d**, The ARPES energy-momentum spectrum shows the static band structure of the heterostructure along $M - K - \Gamma$ and with the ILX at long time delay (100ps).

**Long-lived Floquet-like effects driven by interlayer excitons**

To study the effect induced by a dense population of ILX, we fabricated bilayer heterostructures consisting of 1L $WSe_2$ on 1L $MoS_2$ (Fig. 1b) with twist angles of 2, 8 and 15 degrees. The bilayer heterostructure was stacked on a thin boron nitride (hBN) buffer supported by an n-doped low-resistivity Si substrate, as in previous studies of excitons by TR-ARPES [12,14–17]. The sample was then loaded into a time-of-flight momentum microscope (ToF-MM) for the TR-ARPES measurements. The ToF-MM enables the acquisition of a multidimensional photoemission spectrum $I(E, k_x, k_y)$ from the selected $\sim 10\,\mu m \times 10\,\mu m$ region of the heterostructure. We first characterized the unperturbed band structure of the heterostructure at 100 K. An energy-momentum cut along the Γ–K direction of the $WSe_2$ (Fig. 1d left) shows 450-meV spin-orbit splitting between the first (VB1) and second (VB2) valence bands in the K valley of the $WSe_2$, with the energy scale referenced at the valence band maximum (VBM). The top valence bands of $MoS_2$ appear in the K valley as a weaker and broad peak around -0.9 eV (Fig. 1d), because of the attenuated photoemission intensity from the buried $MoS_2$.

To form dense populations of ILX, we used a 1.72-eV optical pump pulse that resonantly creates an initial dense intralayer exciton population up to $n_X \sim 6.6 \times 10^{12} cm^{-2}$ in $WSe_2$. Following the optical excitation, the intralayer excitons scatter within a picosecond into ILXs, with electrons located the $MoS_2$ layer while the holes remain in the $WSe_2$ layer[15,18–20], providing long population lifetime due to the optically dark nature inherited from the layer- and momentum- separation of the electrons and holes. At a 100 ps time delay, the photoemission spectrum shows an electron signal from ~1.1 eV above the VBM (Fig. 1d right), corresponding to the energy of the ILX as previously reported in TR-ARPES experiments[15,18]. The momentum resolution of our experiment enables direct mapping of the moiré structure, allowing us to determine both the twist angle and the momentum mismatch between the K valleys of the two twisted layers, as shown in Fig. 1c. In the 15-degree-twist heterostructure as discussed in the following sections, the twist creates $0.33\ \text{Å}^{-1}$ momentum offset between the K valleys of the $WSe_2$ and $MoS_2$ layers and therefore the center-of-mass momentum is Q=$0.33\ \text{Å}^{-1}$.

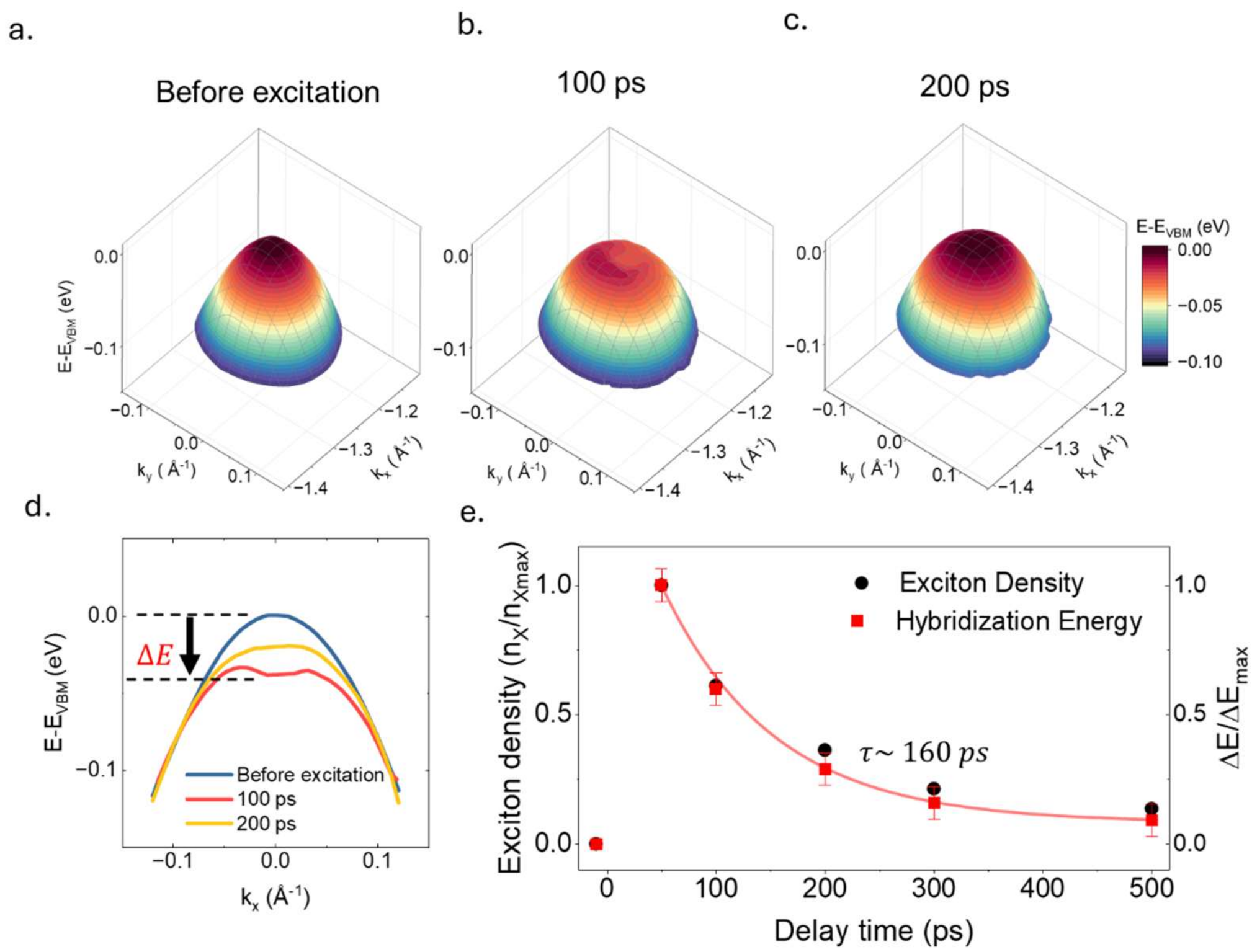

**Figure 2. The long-lived Floquet-like effects with hybridization by interlayer excitons in 15-degree twist heterostructure. a-c**, the energy-momentum dispersion relation of the top valence electrons extracted from experimental data. The dispersion of the top valence band is a parabolic shape without perturbation (a). At 100 ps with dense ILXs (b), the dispersion shows a Mexican-hat-like dispersion and then slowly recover after 200 ps (c). The hybridization persists along with the long-lived high density ILXs. **d**, The 1D energy-momentum dispersion. The hybridization energy $\Delta E$ is extracted from the absolute value of the energy change of the valence band dispersion at the center of K valley. **e**, Long-time dynamics of the ILX population and $\Delta E$, normalized by the maximum density ($6.6 \times 10^{12} cm^{-2}$) and hybridization energy (~70 meV) at 50 ps. The error bars represent the confidence interval while extracting the hybridization energy.

Now, we show that a dense and long-lived incoherent ensemble of ILX can exhibit the same spectral features as seen for coherent exciton-Floquet. We analyzed the temporal evolution of the dispersion of the top valence band (see methods and SI Fig S1) where the holes of the ILX are located. At 100 ps time delay and with a high density of ILX ($\sim 4 \times 10^{12} cm^{-2}$), we clearly see that the unperturbed parabolic dispersion of valence band (Fig.2a) changes to a Mexican-hat-like dispersion (Fig. 2b) – the band curvature opens in upward direction around the valley center and downward at the edges. This effect is similar to the one reported in the exciton-driven Floquet regime, where the coherent exciton coupling effectively hybridizes the dressed conduction band and the valence band within a momentum region corresponding to the envelope function of the exciton resulting in the Mexican-hat dispersion. However, our observation falls beyond the coherent dressing by an exciton field as it persists on timescale far exceeding the typical coherence time of excitons. Here, the coherence of the ILX ensemble is prohibited since they are not formed directly by the coherent laser excitation but instead by complex phonon-assisted scattering processes involving intermediate states at the Q-valley[18,21]. Time-resolved measurements of the ILX population over sub-nanosecond timescales show that the hybridization energy $\Delta E$ decreases as the ILX density decays (Fig. 2e, SI Fig S4). Interestingly, the hybridization energy extracted from the dispersion near the VBM scales linearly with the ILX density (SI Fig. S3c), with a coefficient of $12 \times 10^{-12} meVcm^2$.

**Microscopic picture of the static coupling in the incoherent regime**

Our observation of band hybridization driven by a dense, incoherent ensemble of ILX, together with the linear scaling of the hybridization energy with ILX density, is consistent with the predictions of the time-resolved ARPES theory for coherent and incoherent excitons recently developed in Ref [13]. Within this framework, the coupling between conduction and valence bands induced by a coherent exciton field is captured at the Hartree plus statically screened exchange (HSEX) level. By contrast, the effective static field generated by an incoherent exciton population arises only within a higher-order diagrammatic treatment, such as the T-matrix plus screened exchange (TSEX) approximation.

The dominant process underlying the TSEX description is the conversion of a conduction-band electron with momentum $k$ and energy $\epsilon_{\boldsymbol{k}c}$ into an exciton with centre-of-mass momentum $\boldsymbol{Q}$ and energy $E_X^{\boldsymbol{Q}}$, together with a valence-band electron at momentum $\boldsymbol{k}-\boldsymbol{Q}$ and energy $\epsilon_{\boldsymbol{k}-\boldsymbol{Q}v}$: e($\boldsymbol{k}c$)→X($\boldsymbol{Q}$)+e($\boldsymbol{k}-\boldsymbol{Q}v$). The scattering amplitude $\delta\Sigma_X^{\boldsymbol{Q}}$ for this process scales as $\sqrt{N_X^{\boldsymbol{Q}}}$, where $N_X^{\boldsymbol{Q}}$ is the occupation of the incoherent exciton mode[16]. The electron-removal energy associated with the hybridized conduction-band state is then given by the lowest eigenvalue of the effective 2×2 Hamiltonian $\begin{pmatrix} \epsilon_{\boldsymbol{k}c} & \delta\Sigma_{\boldsymbol{k}X}^{\boldsymbol{Q}} \\ \delta\Sigma_{\boldsymbol{k}X}^{\boldsymbol{Q}*} & E_X^{\boldsymbol{Q}} + \epsilon_{\boldsymbol{k}-\boldsymbol{Q}v} \end{pmatrix}$, that is

$$E_{\boldsymbol{k}}^{\boldsymbol{Q}} = \frac{1}{2}\left( E_X^{\boldsymbol{Q}} + \epsilon_{\boldsymbol{k}-\boldsymbol{Q}v} + \epsilon_{\boldsymbol{k}c} - \sqrt{\left(\epsilon_{\boldsymbol{k}c} - E_X^{\boldsymbol{Q}} - \epsilon_{\boldsymbol{k}-\boldsymbol{Q}v}\right)^2 + 4\left|\delta\Sigma_{\boldsymbol{k}X}^{\boldsymbol{Q}}\right|^2} \right)$$

This expression reveals three key features of the exciton-induced reconstruction. First, for $\boldsymbol{Q}=0$, it reduces to the same hybridization structure obtained for a coherent exciton field within HSEX, upon identifying $\delta\Sigma_X^{\boldsymbol{Q}=0} \to \delta\Sigma_X e^{iE_X t}$ with the corresponding coherent coupling. A zero-momentum incoherent exciton population can therefore generate a Floquet-like band reconstruction without requiring a phase-coherent bosonic field. Second, finite-$\boldsymbol{Q}$ excitons hybridize the conduction band with a replica of the valence band shifted by $\boldsymbol{Q}$ in momentum and by $E_X^{\boldsymbol{Q}}$ in energy, introducing momentum transfer as an additional degree of freedom beyond conventional Floquet engineering. Third, the density-dependent hybridization captures the evolution from a dome-shaped dispersion at low ILX density to a Mexican-hat-like dispersion at high density. As exciton density increases, the valence replica band $E_X^{\boldsymbol{Q}} + \epsilon_{k-\boldsymbol{Q}v}$ undergoes a blueshift and can cross the conduction band $\epsilon_{\boldsymbol{k}c}$, which concomitantly redshifts. This crossing gives rise to strong hybridization between the two bands. The corresponding hybridization energy $\Delta E = E_0^{\boldsymbol{Q}}[N_X^{\boldsymbol{Q}}=0] - E_0^{\boldsymbol{Q}}[N_X^{\boldsymbol{Q}}] \propto N_X^{\boldsymbol{Q}}$ scales linearly with the ILX density, in agreement with Figure 2e.

Here we compare the predictions from the time-resolved ARPES theory[13] with the experiments by examining *finite-momentum* Floquet-like effects with varying the density of the ILXs in heterostructure with 15-degree twist angle and 0.33 $\text{Å}^{-1}$ momentum offset, spanning one order of magnitude from the low density $n_X \sim 4 \times 10^{11}\text{cm}^{-2}$ to the high density $\sim 4 \times 10^{12}$ $\text{cm}^{-2}$ (Fig 3. a-c). At low density limit ($n_X \sim 4 \times 10^{11} cm^{-2}$) as shown in Figure 3.a, a sharp satellite band appears in the photoemission signal ~1.1 eV above VBM, duplicating the dispersion of the downward parabolic valence band around the K point of $WSe_2$. As the ILXs density enters the high-density regime ($n_X \sim 4 \times 10^{12} cm^{-2}$) in Fig. 3c, both the valence band and the excitonic satellite evolve from dome-shaped to Mexican-hat-like dispersions, reflecting strong hybridization between the conduction band and the exciton-dressed valence band. The dispersion of conduction band near minimum is not directly observed in these experiments due to the lack of occupation.

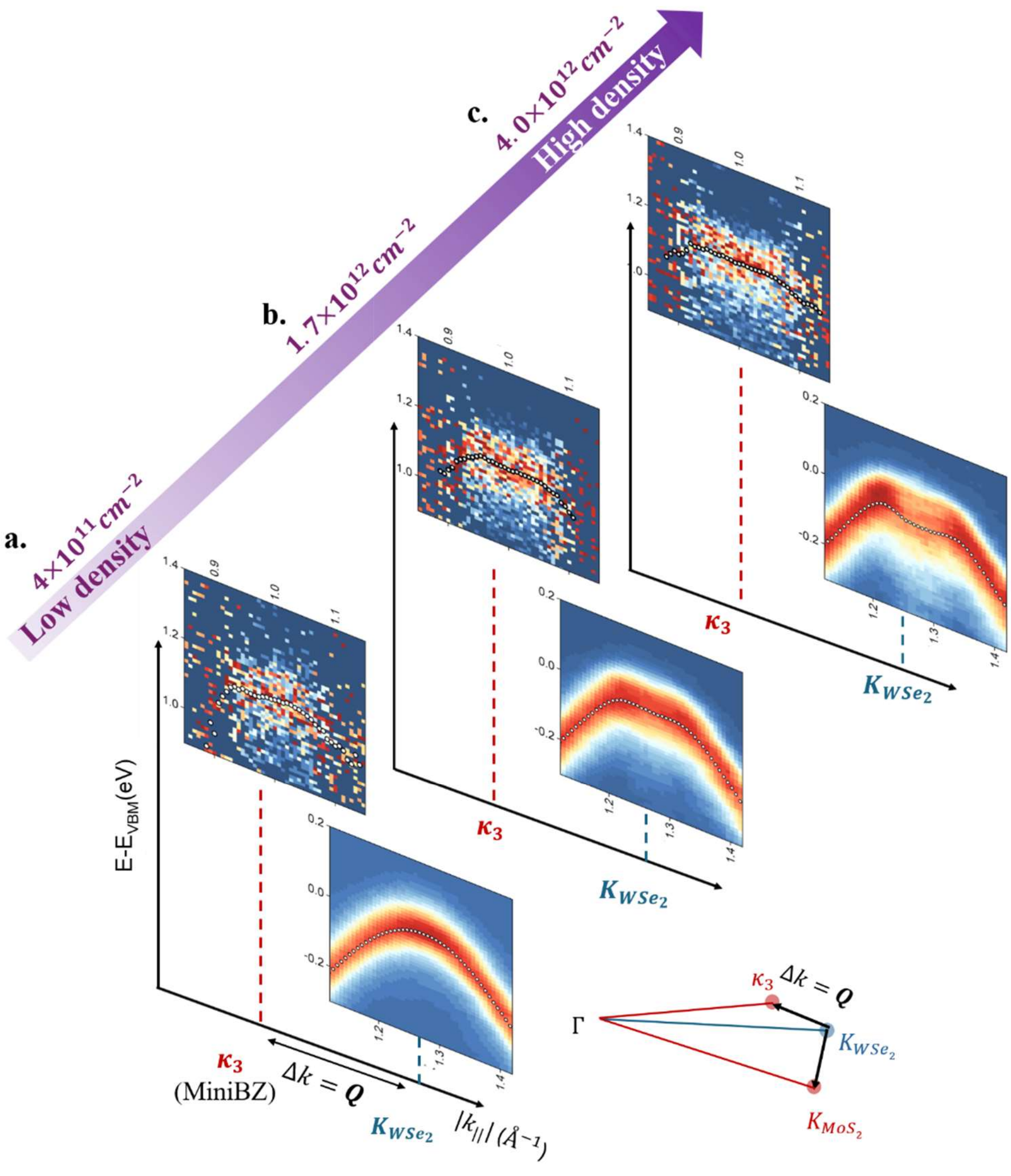

**Figure 3. The effect of hybridization with high-density interlayer excitons. a-c**, the replica (upper panel) and the top VB in K valley (lower panel) with momentum offset in reciprocal space determined by the twist angle (15 degrees). The replica appears at corners of miniBZ and the cut is along $\kappa_3 - \Gamma$ direction. The top VB is shown along $K_{WSe_2} - \Gamma$ direction as shown in the bottom schematics. The dispersion of the valence band changes from a paraboloid shape into Mexican-hat-like dispersion as the exciton density increases (downside panel) and this effect also show up in the replica (upper panel). The density dependence was measured at delay time >50 ps.

## Engineering of the coupling in momentum space

Having established that incoherent excitons can produce Floquet-like effects through a static coupling, we now show how the momentum offset of this coupling can be controlled by adjusting the twist angle. In twisted bilayers, one can manipulate the twist angle of two layers so that the ILXs carry a finite momentum and couple two bands with large momentum offset - the conduction band from $MoS_2$ and the valence band from $WSe_2$, where the electrons and holes are located separately. Here, we artificially stacked the bilayer with different angles - 2, 8 and 15 degrees. As twist angle increases, we can clearly observe the increase in momentum offset between the excitonic satellite bands and the K valley center of $WSe_2$ extracted from VB dispersion with a net momentum of up to 20% size of the BZ (Fig 4 a-c) for a 15° twist angle. This was observed as the signature of ILXs with the moiré potential of the bilayer heterostructure in momentum space, previously shown in tr-ARPES[15,18].

At high ILX densities, we extracted the valence-band dispersion around the VBM and show that it always develops a Mexican-hat-like profile across the investigated twist-angle range (Fig. 4d–f), demonstrating that the ILX-induced static coupling persists up to sub-nanoseconds timescales. The dispersion of the satellite bands duplicates the Mexican-hat-like dispersion of valence bands, with the momentum and energy offset determined by ILX (Fig 4. e-g). These observations provide compelling experimental evidence for one of the most striking consequences of the time-resolved ARPES theory[13]: *incoherent dark excitons can induce static band hybridization at finite momentum transfer*. The accessible momentum offset greatly exceeds that attainable with photons or coherent excitons and, in principle, can be further extended through device engineering.

Furthermore, we would like to note that the nature of the dense and long-lived ILXs also satisfies key ingredients required for the collective phases such as exciton condensate: excitons persist long enough to equilibrate, and the electron and hole kinetic energies are suppressed. Spatially indirect electron-hole system have attractive intense research to explore the collective phenomenon of excitons[22,23]. The TMDC bilayers further benefit from the strong exciton interactions that hold the promise for high-temperature condensates[24,25]. Our experiments, by optical injection of carriers, directly observe the dynamic change of electronic structure, offering a clean platform and a controllable manner to explore the underlying physics and novel phases.

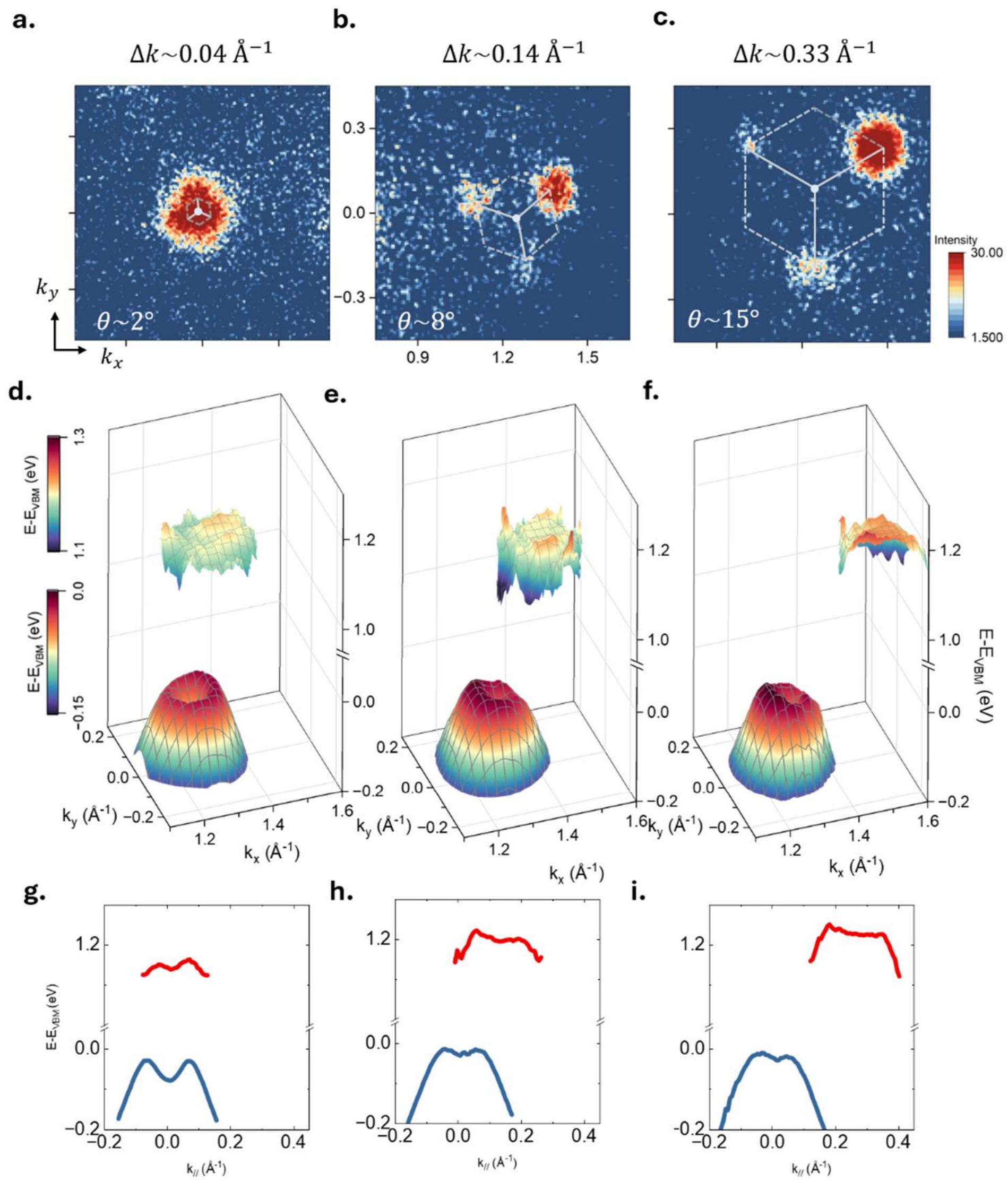

**Figure 4. Hybridization in the valence band and its replica with finite momentum offset. a-c,** the miniBZ created by the twist between $MoS_2$ and $WSe_2$ at 2-, 8-and 15-degree twist. The replicas appear at the corners of mimiBZ. **d-f.** the 3D dispersion extracted for the top valence band in K valley and the replica with exciton density around $3 \times 10^{12} cm^{-2}$, for the 2-degree twist heterostructure (d), For the 8-degree twist heterostructure (e), and for the 15-degree twist heterostructure (f). **g-i,** the corresponding 1D cut of the energy-momentum distribution of the valence electron and the replica for three heterostructures at different twist angles.

## Conclusion and prospects

Our work established a powerful approach to achieve Floquet-like effects using incoherent excitons. The strong hybridization, long-lived nature, robustness against scattering, overcomes the challenges that bottleneck further applications of Floquet physics. Previous efforts to achieve Floquet engineering, particularly for the quasi-equilibrium Floquet state[26,27], are mostly limited to optical drive at low frequency (microwave to mid IR range) to take the advantage of stronger light-matter coupling and to avoid the detrimental effects such as heating and carrier absorption, but this strategy is only suitable for gapless or narrow-gap materials. Our work demonstrated quasi-equilibrium band structure engineering that might trigger new correlated electronic states[32–34] in semiconductors with driving frequency compatible with optoelectronic devices. The twisted TMDC heterostructure is one of the favorable systems that one can flexibly tune the energy landscape, momentum offset and moiré confinement to engineer coupling in energy-momentum space. The integration with dark excitons intrinsically enables possibility of combining with other degrees of freedom, including spin, momentum, and orbitals of the bands, leading to the emerging topological nontrivial bands and exotic quasiparticles[10].

## Methods

### Sample preparation

The 1L-$WSe_2$ and 1L- $MoS_2$ flakes were exfoliated from the bulk crystal (hqGraphene) using the polydimethylsiloxane stamps PDMS (gel-pak). A flat and thin hexagonal boron nitride (hBN) flake was directly cleaved on the n-doped Si substrate before stacking the heterobilayer. The selected 1L- $MoS_2$ was stacked onto the hBN area and then immediately the sample is rinsed in acetone and isopropanol followed by annealing at 350 °C for 11 h in the ultra-high vacuum. The 1L-$WSe_2$ was further stacked onto the top of 1L- $MoS_2$ with a relative rotation angle controlled by motor on the transfer stage. The sample was followed by the same cleaning procedure of rinse and annealing. During the transfer process, the crystallographic alignment of 1L-$WSe_2$ and 1L-$MoS_2$ was estimated by the sharp edges of the flakes.

### Experimental details of TR-ARPES

TR-ARPES was performed in a time-of-flight momentum microscope (ToF MM) (Metis 1000, SPECS GmbH), which was connected to the home-built table-top high harmonic generation (HHG) setup. The HHG setup was driven by a Yb-doped fiber laser system operating at 2 MHz with 230 fs pulse width. For the optical probe path, out of 50 μJ pulse energy delivered by the laser system out, 35 μJ (1030 nm) was frequency doubled by a thin BBO to generate ~15 μJ of 515 nm and was then focused onto a Kr gas jet to generate the XUV harmonics. The Sn and Al filters were used to select the XUV probe at 21.7 eV. The XUV probe was further collected by ellipsoidal mirror and focused onto the sample in the momentum microscope. The rest of the power (15 $\mu J$ 1030 nm) from the laser was delivered to the noncollinear optical parametric amplifier (NOPA) to convert to tunable light pulse in the vis-NIR region. Then the pump light pulse was focused onto sample. To efficiently generate intralayer exciton in $WSe_2$, the pump photon energy was set at 1.72 eV.

The photoemitted electrons were further collected by an immersion objective lens and analyzed by the ToF MM. A field aperture inserted at the Gaussian image plane of the electron optics was used to select the electrons emitted from a $10 \times 10\ um^2$ region of heterostructure. The momentum distribution of electrons was further imaged from the back focal plane of the objective lens, and the energy-resolved spectrum are obtained by the time-of-flight analyzer. All TR-ARPES measurements were conducted with sample temperature at 100 K.

### Dispersion analysis of the valence bands

Based on the 3D dataset $I(k_x, k_y, E)$ from ToF-MM, we further analyzed the valence band dispersion and tracked the changes due to the presence of interlayer excitons. A small cubic region ($k_\parallel - K = \pm 0.15$ Å$^{-1}$; $E = +0.5\ eV\ to\ -1.2\ eV$) was chosen around the top of the valence band in K valley. For each $k_{||}$, we obtained the energy distribution curve (EDC) by integrating in the neighboring region $k_\parallel = \pm 0.024$ Å$^{-1}$ to improve the signal-noise ratio. To further extract the dispersion relation of valence bands, a Shirley-type background was subtracted from the EDC then a multi-Gaussian function was used to fit the three major peaks together. The first two peaks (VB1 and VB2) are mainly contributed by the spin-split VB from $WSe_2$, and the third peak is composed of the valence bands of $MoS_2$. This EDC fitting process was repeated for each $k_{||}$ in the range of $k_{||} = \pm 0.15$ Å$^{-1}$ in the K valley so we can map out the valence band dispersion $E_{VB}(k_x, k_y)$.

This process was also repeated for all the different photoexcitation densities shown in Fig. 2-4. In the 2D image, momentum-energy cuts along Γ–K–M high-symmetry points of BZ of $WSe_2$ were chosen to show the momentum-energy distribution of photoemission counts and the extracted dispersions from the top valence band.

**Dispersion analysis of valence band replica**

The valence band replica appeared at ~1.1 eV and the corners of miniBZ. Similarly, we chose the 3D dataset around the corners of miniBZ ($k_{\parallel} - \kappa = \pm 0.15$ Å$^{-1}$; $E = +1.5\ eV\ to + 0.7\ eV$). For each $k_{||}$ position around the K point (±0.2 Å−1), we obtained the EDC by integrating in the neighboring region $k_{\parallel} = \pm 0.04$ Å$^{-1}$. The EDC was further fitted by a single Gaussian function to obtain the dispersion as a function of $k_{||}$.

**Calculating the magnitude of hybridization.**

After obtaining the valence band dispersion for different ILX densities, we matched the VB1 and VB2 dispersions for the unphotoexcited and photoexcited cases away from the K point ($k_{||} < -0.15$ Å$^{-1}$). This step removes the energy shift produced by surface photovoltage effect. We then subtracted the photoexcited VB1 dispersion from the unphotoexcited VB1 dispersion to get the $\Delta E_X$ curve for different exciton densities. We clearly observed a substantial change in the VB1 dispersion with increasing exciton density. No change was observed in the dispersion for VB2, which is consistent with the prediction that only a limited section of VB1 was affected by incoherent static coupling produced by ILXs and the rest of the band structure was left unchanged. To calculate the magnitude of $\Delta E_X$, we used a Gaussian function to approximate the change of dispersion at each $k_{||}$ and extract the peak value.

**Extraction of the exciton density from the hole occupation**

From TR-ARPES data, we calculated the absolute exciton density from the hole occupation as previous studies[12,15]. We focused the analysis on the top of the valence band VB1 only where the holes of ILXs are located. We compared the photoemission intensity between the unphotoexcited and photoexcited band structures for VB1. First, we can precisely locate the VBM and remove any momentum and energy offset between datasets by fitting the valence band dispersion $E_{VB1}(k_x, k_y)$ with a 2D parabolic function $E_{VB1} = a(k_{||} - K)^2 + E_{offset}$.Then we can obtain the intensity distribution of VB1 in the K valley for both photoexcited $I_{photoexcit}\ \ [k_x, k_y]$ and unphotoexcited $I_{unphotoexcited}[k_x, k_y]$. Next, we need to eliminate the intensity variation between the measurements due to experimental conditions. For this, the data were normalized by the intensity in a small k-space region around $[k_{x0}, k_{y0}]$ which is away from the K valley center and unaffected by the presence of hole. The occupation distribution of holes in the 2D momentum space can be represented as:

$$f_{hole}[k_x, k_y] = 1 - \frac{I^{Ref}_{unphotoexicted} I_{photoexicted}[k_x, k_y]}{I^{Ref}_{photoexcited} I_{unphotoexc}\ \ \ \ [k_x, k_y]}$$

Finally, the hole density was given by multiplying the hole occupancy distribution with the density of hole states in momentum space and summing over the entire distribution in momentum space:

$$n_{hol} \; = 2/L^2 \iint f_{hole}[k_x, k_y] dk_x dk_y / (4\pi/L^2)$$

where $dk_{x,y} = 0.0066$ Å$^{-1}$ is the k-space coordinate spacing in the $k_{x,y}$ directions. The prefactor 2 is for considering the two K valleys in the BZ. $L^2$ is the volume of the 2D system.

**Acknowledgement**

X.Z., J. H., V.P., D.R.B., H.S., N. T., A. M., T.F., S.R., J. F., M.K.L.M., J.M. and K.M.D. acknowledge that this work was supported by the Femtosecond Spectroscopy Unit of OIST. K.M.D. acknowledges support from the JST FOREST (Grant number 23718777) and the JSPS KAKENHI (Grant numbers 22K18270, 24H00191 and 23K25807). J.M. acknowledges support from the JSPS KAKENHI (Grant Number 24K00561). T.F. acknowledges support from the JSPS KAKENHI (Grant Number 25K17332). E.P. and G.S. acknowledge funding from INFN through project TIME2QUEST and from the Horizon Europe research and innovation program of the European Union under the Marie Sklodowska-Curie Grant Agreement No. 101118915 (TIMES). K.W. and T.T. acknowledge support from the CREST (JPMJCR24A5), JST and World Premier International Research Center Initiative (WPI), MEXT, Japan.

**Author contributions**

X.Z., V.P., D.R.B., J.H performed the TR-ARPES measurements with the assistance of N.T., H.S., T.F, J.M. and M.K.L.M. X.Z., J.H. performed analysis of the data with assistance of J.M. and M.K.L.M. S.R., X.Z., O.K. and J.F. fabricated and characterized the samples. K.W. and T.T. provided the bulk hBN crystals for the study. E.P. and G.S. performed the theoretical calculations. Project was supervised by K.M.D. All authors contributed to the writing and reviewing of the manuscript.

**Competing interests**

J.M., M.K.L.M. and K.M.D. are inventors on a granted patent related to this work filed by the Okinawa Institute of Science and Technology School Corporation (US patent 11,372,199). The authors declare no other competing interests.

**Data availability**

The data that supports the findings of this work are available upon request to the corresponding author.

## Supplementary Information

1. **Determination of the twist angle of the bilayer heterostructure**
2. **Dispersion analysis for the band structure of the valence band with and without ILX**
3. **Hybridization energy extracted from heterostructures with different twist angles**
4. **Long-lived effect at sub-nanosecond**

### 1. Determination of the twist angle of the bilayer heterostructure

We first estimate the twist angle from the sharp edge of the monolayer crystal. This angle is further confirmed by the size of miniBZ, obtained by the K valley center offset from the VBM of $WSe_2$ and $MoS_2$ before photoexcitation, or the weighted center of ILX-bound electrons and intralayer exciton-bound electrons after photoexcitation. In the bilayer heterostructure with twist angle (2， 8 or 15 degrees), the 2D momentum distribution of the excited states by integrating signals around 1.1 eV exhibits a clear momentum offset relative to the orientation of $WSe_2$ (Fig. 1c). The corners of the miniBZ are brightened, as determined by the twist angle and the 3% lattice mismatch between the $WSe_2$ and $MoS_2$.

### 2. Dispersion analysis for the band structure of the valence band with and without ILX

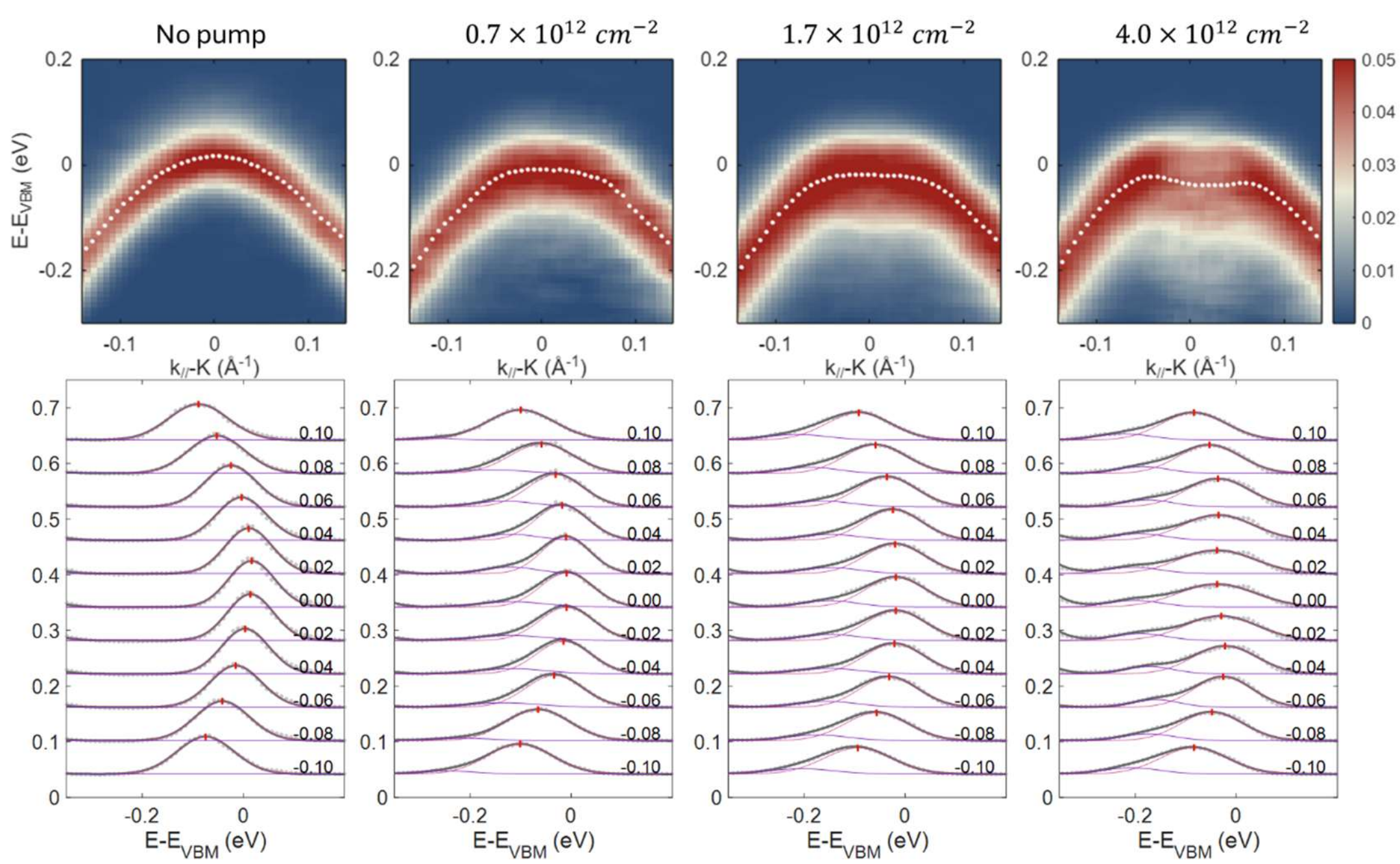

**S1. 15-degree twist heterostructure**. Energy distribution curves at different $k_{//}$ values relative to the K valley center are normalized by the area under curve. Two-Gaussian function is used to fit the EDC of VB1. The dispersion (red mark on top of first Gauss) was extracted from the major peak of the first Gaussian. The second Gauss (purple line) picks the tail part.

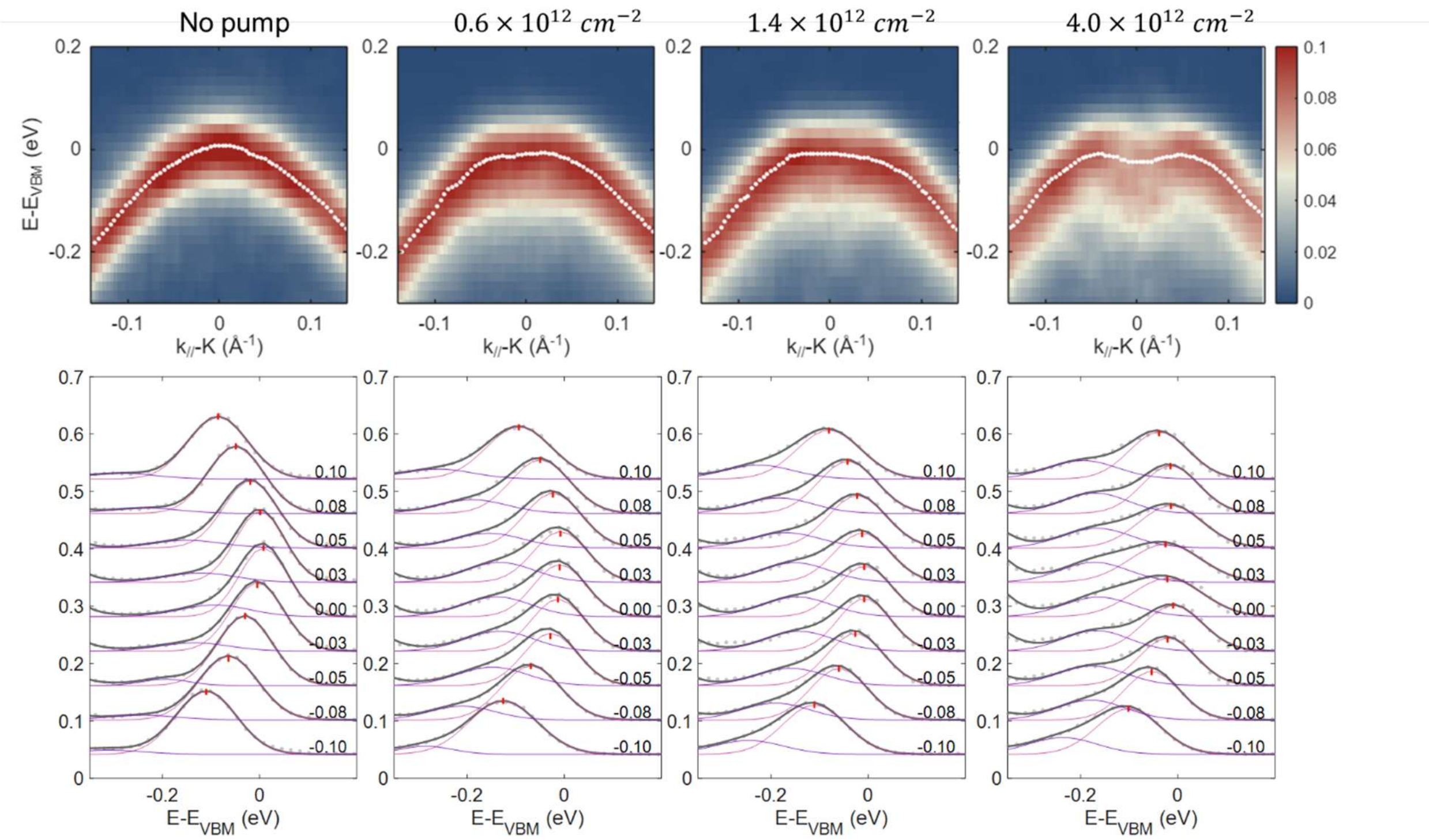

**S2. 2-degree twist heterostructure.** Energy distribution curves at different $k_{//}$ values relative to the K valley center are normalized by the area under curve.

### 3. Hybridization energy extracted from heterostructures with different twist angles

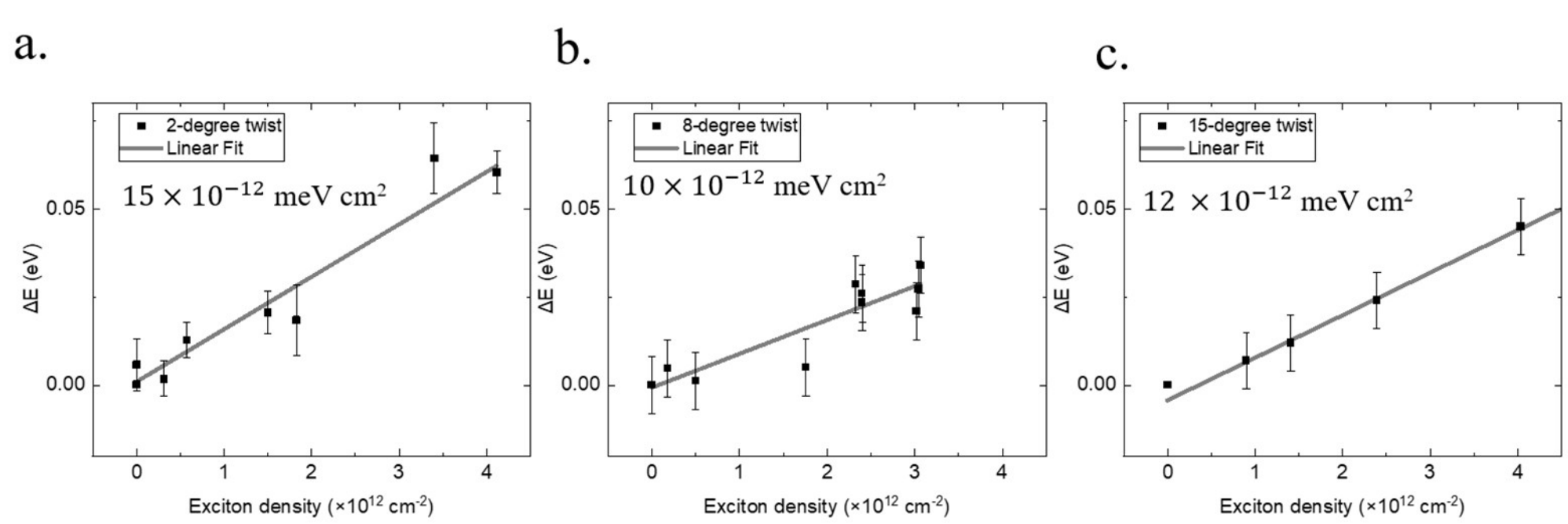

**S3. Hybridization energy versus the density of ILXs in heterostructure with different twist angles.** The error bar of hybridization energy is extracted from 95% confidence interval of EDC fitting.

4. **Long-lived effect at sub-nanosecond**

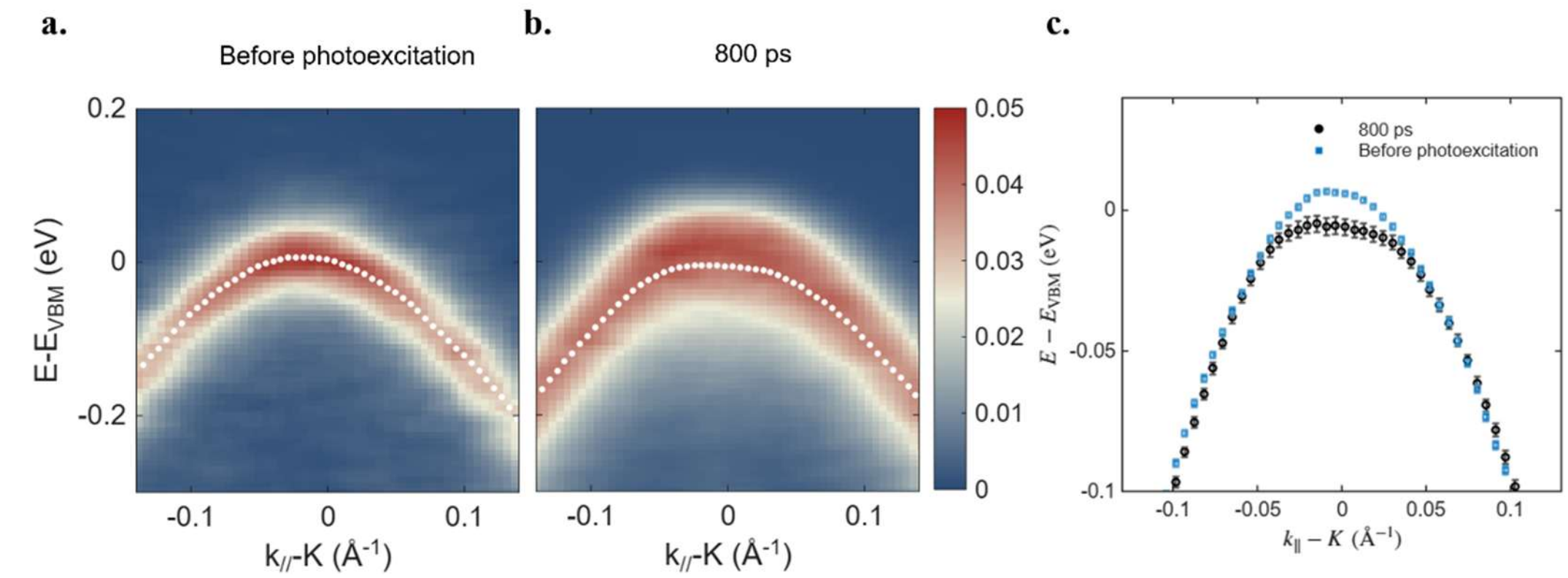

**S4. The valence band dispersion with ILX in 8-degree twist heterostructure at 800 ps after photoexcitation.** a. the valence band dispersion before photoexcitation as reference. b. the valence band dispersion when the exciton density at 800 ps is $2.4 \times 10^{12}\ cm^{-2}$. c. the valence band dispersion with the 95% confidence interval of EDC fitting at each $k_{||}$.